\documentclass[pra,aps,preprint,superscriptaddress,amsmath,amssymb]{revtex4-1}
\usepackage{graphicx}
\usepackage{multirow}
\usepackage{color}
\usepackage{ulem}
\usepackage{comment}

\newcommand{\ket}[1] {\left| #1 \right\rangle}

\begin{document}

\title{Inversion of Qubit Energy Levels in Qubit-Oscillator Circuits in the Deep-Strong-Coupling Regime}

\author{F. Yoshihara}
\email{fumiki@nict.go.jp}
\affiliation{Advanced ICT Institute, National Institute of Information and Communications Technology, 4-2-1, Nukuikitamachi, Koganei, Tokyo 184-8795, Japan}
\author{T. Fuse}
\affiliation{Advanced ICT Institute, National Institute of Information and Communications Technology, 4-2-1, Nukuikitamachi, Koganei, Tokyo 184-8795, Japan}
\author{Z. Ao}
\affiliation{Department of Applied Physics, Waseda University, 3-4-1, Ookubo, Shinjuku-ku, Tokyo 169-8555, Japan}
\author{S. Ashhab}
\affiliation{Qatar Environment and Energy Research Institute, Hamad Bin Khalifa University, Qatar Foundation, Doha, Qatar}
\author{K. Kakuyanagi}
\affiliation{NTT Basic Research Laboratories, NTT Corporation, 3-1 Morinosato-Wakamiya, Atsugi, Kanagawa 243-0198, Japan}
\author{S. Saito}
\affiliation{NTT Basic Research Laboratories, NTT Corporation, 3-1 Morinosato-Wakamiya, Atsugi, Kanagawa 243-0198, Japan}
\author{T. Aoki}
\affiliation{Department of Applied Physics, Waseda University, 3-4-1, Ookubo, Shinjuku-ku, Tokyo 169-8555, Japan}
\author{K. Koshino}
\affiliation{College of Liberal Arts and Sciences, Tokyo Medical and Dental University, 2-8-30, Kounodai, Ichikawa, Chiba 272-0827, Japan.}
\author{K. Semba}
\email{semba@nict.go.jp}
\affiliation{Advanced ICT Institute, National Institute of Information and Communications Technology, 4-2-1, Nukuikitamachi, Koganei, Tokyo 184-8795, Japan}

\date{\today}

\begin{abstract}
We report on experimentally measured light shifts of superconducting flux qubits deep-strongly coupled to
LC oscillators,
where the coupling constants are comparable to the
qubit and oscillator resonance frequencies.
By using two-tone spectroscopy, the energies of the six lowest levels of each circuit are determined.
We find huge Lamb shifts that exceed 90\% of the bare qubit frequencies and
inversions of the qubits' ground and excited states when there
are a finite number of photons in the oscillator.
Our experimental results agree with theoretical predictions based on the quantum Rabi model.
\end{abstract}

%PACS numbers:

\maketitle
%introduction
According to quantum theory, the vacuum electromagnetic field has ``half-photon" fluctuations,
which cause several physical phenomena such as the Lamb shift~\cite{Lamb47PR}.
A cavity can enhance the interaction between the atom and the electromagnetic field inside the cavity
and enables more precise measurements of the influence of the vacuum.
Cavity/circuit-quantum-electrodynamics systems are usually well described by the Jaynes-Cummings Hamiltonian~\cite{Blais04PRA,walls2007quantum}.
In the strong-coupling regime, when the cavity's resonance frequency $\omega$ is on resonance with the atom's transition frequency $\Delta$, the vacuum Rabi splitting~\cite{Thompson92PRL,Wallraff04Nat,Kato15PRL} and oscillations~\cite{Brune96PRL,Johansson06PRL} have been observed.
In the off-resonance case, the Lamb shift~\cite{Heinzen87PRL,Brune94PRL,Fragner08Science} caused by the vacuum fluctuations and the ac-Stark shift proportional to the photon number in the cavity were observed~\cite{Brune94PRL,Schuster05PRL,Schuster07Nature,Fragner08Science}.
In the so-called ultrastrong-coupling regime~\cite{Niemczyk10NatP,Forn10PRL}, where the coupling constant $g$ becomes around 10\% of $\Delta$ and $\omega$, and the deep-strong-coupling regime~\cite{Yoshihara17NatPhys,Yoshihara17PRA},
where $g$ is comparable to or larger than $\Delta$ and $\omega$,
the rotating-wave approximation used in the Jaynes-Cummings Hamiltonian breaks down, and the system should be described by the quantum Rabi Hamiltonian~\cite{Rabi37PR, Jaynes63IEEE, Braak11PRL}.
In these regimes, the light shifts of an atom could nonmonotonically change as $g$ increases, and the amount of the shift is not proportional to the photon number in the cavity~\cite{Ashhab10PRA,Rossatto17PRA}.

\begin{figure}
\includegraphics{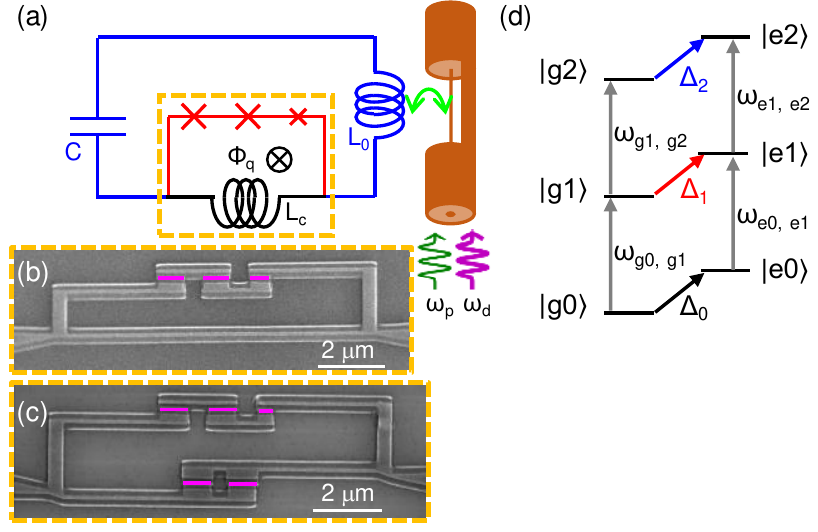}
\caption{(a) Circuit diagram.
A superconducting flux qubit (red and black) and a superconducting LC oscillator (blue and black) are inductively coupled to each other by sharing an inductance (black).
(b), (c) Scanning microscope images of the qubit and the shared inductance located at the orange rectangle in diagram (a).
Josephson junctions are represented by magenta rectangles.
The shared inductance is a superconducting lead (b) or a loop of Josephson junctions (c).
(d) The diagram of the six lowest energy levels of a qubit-oscillator circuit.
The energy eigenstates are expressed as $|in\rangle$ $(i = \textrm{g},\,\textrm{e}$ and $n = 0,$ 1, 2, $\cdots$),
which indicates that the qubit is in ``g'' the ground or ``e'' the excited state and
the number of real photons in the oscillator is $n$.
The arrows indicate transition frequencies between energy eigenstates
and also mean that the transitions are allowed.
Here, $\Delta_n$ ($n=0$, 1, 2) is the photon-number-dependent qubit frequency.
}
\label{Fig:CircuitElevels}
\end{figure}

%intro 2nd part
In this work, to study the light shift in the case of $g\sim \omega$,
we investigated qubit-oscillator circuits that each comprise a superconducting flux qubit~\cite{Mooij99Sci} and an LC oscillator inductively coupled to each other by sharing a loop of Josephson junctions that serves as a coupler [Figs.~\ref{Fig:CircuitElevels}(a) and \ref{Fig:CircuitElevels}(c)].
By using two-tone spectroscopy~\cite{Fink08Nature,Abdu10PRL}, the energies of the six lowest energy eigenstates were measured, and the photon-number-dependent qubit frequencies were evaluated.
We find Lamb shifts over 90\% of the bare qubit frequency and inversions of the qubit's ground and excited states when there are a finite number of photons in the oscillator.

%figure circuit diagram and energy levels

%circuit and Hamiltonian
The qubit-oscillator circuit is described by the Hamiltonian
\begin{equation}
\hat{H} = -\frac{\hbar}{2}(\Delta\hat{\sigma}_x + \varepsilon\hat{\sigma}_z) +\hbar\omega \hat{a}^{\dagger}\hat{a}+ \hbar g \hat{\sigma}_z \left( \hat{a} + \hat{a}^{\dagger} \right).
\label{Eq:Hamiltonian}
\end{equation}
The first two terms represent the energy of the flux qubit written in the basis of two states with persistent currents flowing in opposite directions around the qubit loop, $\left |\circlearrowleft \right \rangle_{\rm q}$ and $\left |\circlearrowright \right \rangle_{\rm q}$.
The operators $\hat{\sigma}_{x,z}$ are the standard Pauli operators.
The parameters $\hbar \Delta$ and $\hbar \varepsilon$ are the tunnel splitting and the energy bias between $\left |\circlearrowleft \right\rangle_{\rm q}$ and $\left |\circlearrowright \right \rangle_{\rm q}$,
where $\hbar \varepsilon$ can be controlled by the flux bias through the qubit loop $\Phi_{\rm q}$.
The third term represents the energy of the LC oscillator, where $\omega = 1/\sqrt{(L_0 + L_{\rm c})C}$ [see Fig.~\ref{Fig:CircuitElevels}(a)] is the resonance frequency, and $\hat{a}^\dagger$ and $\hat{a}$ are the creation and annihilation operators, respectively.
The fourth term represents the coupling energy.

%quantum Rabi model Hamiltonian
At $\varepsilon = 0$, the Hamiltonian in Eq.~(\ref{Eq:Hamiltonian}) reduces to that of the quantum Rabi model $\hat{H}_{\rm Rabi}$.
In the limit $\Delta \ll \omega$,
the energy eigenstates are well described by Schr\"odinger-cat-like entangled states between persistent-current states of the qubit and displaced Fock states of the oscillator $\hat{D}(\pm \alpha)|n\rangle_{\rm o}$~\cite{Ashhab10PRA,Rossatto17PRA}:
\begin{eqnarray}
\nonumber
|\textrm{g}n\rangle \simeq \frac{\left |\circlearrowleft \right\rangle_{\rm q}\otimes\hat{D} (-\frac{g}{\omega} ) |n\rangle_{\rm o} + \left |\circlearrowright \right\rangle_{\rm q}\otimes\hat{D} ( \frac{g}{\omega} )|n\rangle_{\rm o}}{\sqrt{2}}, 
\label{Eq:gn}
\end{eqnarray}
\begin{eqnarray}
|\textrm{e}n\rangle \simeq \frac{\left |\circlearrowleft \right\rangle_{\rm q}\otimes\hat{D} (-\frac{g}{\omega} ) |n\rangle_{\rm o} - \left |\circlearrowright \right\rangle_{\rm q}\otimes\hat{D} ( \frac{g}{\omega} )|n\rangle_{\rm o}}{\sqrt{2}}. 
\label{Eq:en}
\end{eqnarray}
Here,
$\hat{D}(\alpha) = \exp (\alpha\hat{a}^\dagger - \alpha^*\hat{a})$ is the displacement operator,
and $\alpha$ is the amount of the displacement.
The energy eigenstates on the left-hand side are expressed as $|in\rangle$ ($i$ = g, e), where ``g" and ``e" denote, respectively, the ground and excited states of the qubit and $n$ the number of real photons in the oscillator.
On the right-hand side, $|n\rangle_{\rm o}$ denotes the oscillator's $n$-photon Fock state.
Note that the displaced vacuum state $\hat{D}(\alpha)|0\rangle_{\rm o}$ is the coherent state
$|\alpha\rangle_{\rm o} = \exp (-|\alpha|^2/2)\sum_{n = 0}^{\infty}\alpha^n|n\rangle_{\rm o}/\sqrt{n}$.

The photon-number-dependent qubit frequency $\Delta_n(g/\omega) \equiv \omega_{\textrm{e}n} - \omega_{\textrm{g}n}$ is defined as the energy difference between the energy eigenstates $|\textrm{g}n\rangle$ and $|\textrm{e}n\rangle$, and
it can be expressed as (see the solid lines in Fig.~\ref{Fig:GauLag_both}):
\begin{eqnarray}
\nonumber
\Delta_n(g/\omega) & = & \langle \textrm{e}n|\hat{H}_{\rm Rabi}|\textrm{e}n\rangle - \langle \textrm{g}n|\hat{H}_{\rm Rabi}|\textrm{g}n\rangle \\
\nonumber
& \simeq & \Delta [
_{\rm o}\langle n| \hat{D}^{\dagger}(-g/\omega)\hat{D}(g/\omega) |n\rangle_{\rm o}] \\
& = & \Delta \exp(-2g^2/\omega^2)L_n(4g^2/\omega^2).
\label{Eq:Dn}
\end{eqnarray}
Here, $L_n$ is a Laguerre polynomial:
$L_0(x) = 1$, $L_1(x) = 1-x$, $L_2(x) = (x^2-4x+2)/2$, and so on.
The difference between $\Delta_n$ and the bare qubit frequency $\Delta$ can be considered as the $n$-photon ac-Stark shift $|\Delta_n - \Delta|$.
In particular, $|\Delta_0-\Delta|$ is referred to as the Lamb shift.
Note that the Bloch-Siegert shift~\cite{Bloch40PR,TannoudjiAPinter}, the contribution from the counterrotating terms, is included in the $n$-photon ac-Stark shifts.
Since $L_0 = 1$, a considerable Lamb shift is expected when $g$ becomes comparable to $\omega$.
A similar suppression of transition frequencies because of coupling to other degrees of freedom is well known in polaron physics and other fields.
For example, such an effect was recently discussed for an Andreev-level qubit~\cite{Zazunov05PRB}.
Considering that $L_n$ has $n$ zeros, i.e., points where $L_n(x)$ is equal to zero,
$\Delta_n(x)$ also has $n$ zeros,
and, hence,
in general alternates between positive and negative values.
In other words, the qubit's ground and excited states exchange their roles every time when $\Delta_n = 0$.
The bare qubit frequency $\Delta$ is the tunnel energy between the states $\left | \circlearrowleft \right \rangle_{\rm q}$ and $\left | \circlearrowright \right \rangle_{\rm q}$.
Taking either one of these two states and a finite value of $g$, the oscillator is populated by virtual photons even in the ground state, and the virtual photon states for the qubit states $\left | \circlearrowleft \right \rangle_{\rm q}$ and $\left | \circlearrowright \right \rangle_{\rm q}$ are different from each other.
As a result, the qubit has to ``drag" the oscillator every time it flips its state~\cite{Ashhab10PRA}, which can be seen as an effective reduction of $\Delta$ by a factor that is determined by the overlap integral between the interaction-caused displaced $n$-photon Fock states of the oscillator~\cite{SM} as described by the second line of Eq.~(\ref{Eq:Dn}).
One way to understand negative values of $\Delta_n$ is to think of them as describing a situation where the antibonding state of $\left | \circlearrowleft \right \rangle_{\rm q}$ and $\left | \circlearrowright \right \rangle_{\rm q}$ is more stable than the bonding state.
Note that here the displaced states $\hat{D}(\pm g/\omega)|0\rangle_{\rm o}$ contain only virtual photons, while the states $\hat{D}(\pm g/\omega)|n\rangle_{\rm o}$ for $n \ge1$ contain a mixture of real and virtual photons.

Although Eqs.~(\ref{Eq:en}) and (\ref{Eq:Dn}) are not exact for general values of the circuit parameters, they remain reasonably good approximations as long as $\Delta < \omega$.
Furthermore, the symmetry of $\hat{H}_{\rm Rabi}$ is independent of the circuit parameters, which means that certain transitions will remain forbidden even if the corresponding states do not have simple forms.
These two considerations allow us to easily identify the energies of the different eigenstates from the experimental spectra~\cite{SM}.

\begin{figure}
\includegraphics{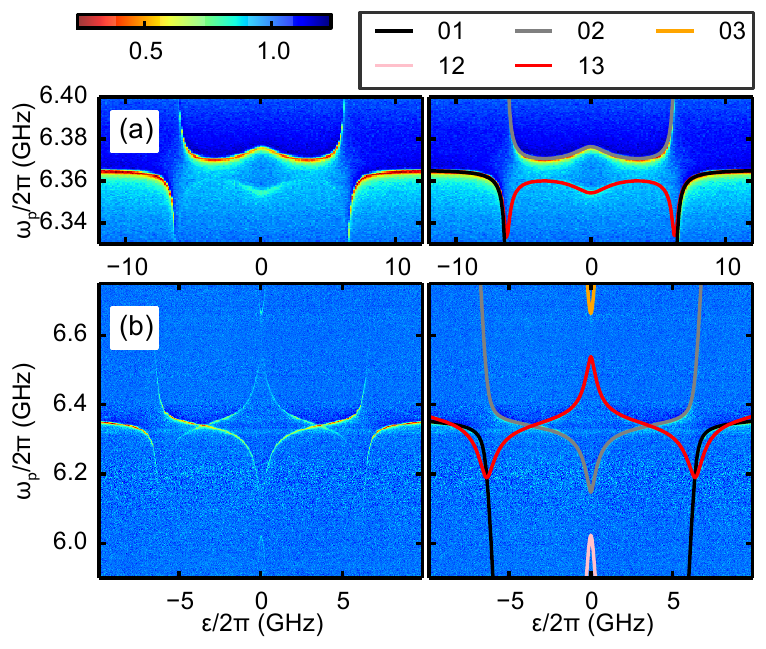}
\caption{Measured transmission spectra for two qubit-oscillator circuits
as functions of the qubit's energy bias $\varepsilon$ and probe frequency $\omega_{\rm p}$.
The color scheme is chosen such that the lowest point in each spectrum is red and the highest point is blue.
The right panels show the transition frequencies calculated from the Hamiltonian to fit experimental data.
The black, gray, orange, pink, and red lines correspond to the transitions $\ket{0}\rightarrow\ket{1}$, $\ket{0}\rightarrow\ket{2}$, $\ket{0}\rightarrow\ket{3}$, $\ket{1}\rightarrow\ket{2}$, and $\ket{1}\rightarrow\ket{3}$, respectively.
The parameters are obtained as
(a)~$\Delta/2\pi=1.246$~GHz, $\omega/2\pi=6.365$~GHz, and $g/2\pi=0.42$~GHz corresponding to set~A; (b)~$\Delta/2\pi=1.68$~GHz, $\omega/2\pi=6.345$~GHz, and $g/2\pi=7.27$~GHz corresponding to set~H.
}
\label{Fig:sparaB}
\end{figure}

%spectroscopy
To determine the parameters of the qubit-oscillator circuits ($\Delta$, $\omega$, and $g$),
spectroscopy was performed by measuring the transmission spectrum through the transmission line that is inductively coupled to the LC oscillator [Fig.~\ref{Fig:CircuitElevels}(a)].
In total, nine sets of parameters (A--I in Table~\ref{Tab:params}) in five circuits were evaluated.
The shared inductance of the circuit for set~A is a superconducting lead
[Fig.~\ref{Fig:CircuitElevels}(b)],
while that of the circuits for sets~B--I is a loop of Josephson junctions
[Fig.~\ref{Fig:CircuitElevels}(c)],
where eight flux bias points in four circuits were used~\cite{SM}.
Therefore, much larger $g$ is expected for sets~B--I.
When the frequency of the probe signal $\omega_{\rm p}$ matches the frequency $\omega_{kl}$ of a transition $|k\rangle \to |l\rangle$,
where $|0\rangle$ stands for the ground state and $|k\rangle$ with $k \ge 1$
stands for the $k$th excited state of the coupled circuit,
the transmission amplitude decreases, provided that the transition matrix element
$\langle k|(\hat{a} + \hat{a}^\dagger)|l\rangle$ is not 0.
Note that because $\varepsilon$ is now generally nonzero,
we have labeled the energy eigenstates using a single integer $k$ instead of the label $|in\rangle$ used above.
Figure~\ref{Fig:sparaB} shows the amplitudes of the transmission spectra
$|S_{21}^{\rm meas}(\varepsilon,\omega_{\rm p})/S_{21}^{\rm bg}(\omega_{\rm p})|$
for sets~A and H.
Here, $\omega_{\rm p}$ is the probe frequency,
and $S_{21}^{\rm meas}(\varepsilon,\omega_{\rm p})$ and $S_{21}^{\rm bg}(\omega_{\rm p})$ are, respectively, the measured and background transmission coefficients~\cite{SM}.

%determine parameters
The circuit parameters are obtained from fitting the experimentally measured resonance frequencies to those numerically calculated by diagonalizing $\hat{H}$ with $\Delta$, $\omega$ and $g$ treated as fitting parameters.
In Fig.~\ref{Fig:sparaB}, the right panels show the calculated transition frequencies superimposed on the measured spectra.
In Fig.~\ref{Fig:sparaB}(a),
one can see the characteristic spectrum of the strong- and ultrastrong-coupling regimes.
From the fitting, the parameters are obtained as
$\Delta/2\pi=1.246$~GHz, $\omega/2\pi=6.365$~GHz, and $g/2\pi=0.42$~GHz.
The spectrum shown in Fig.~\ref{Fig:sparaB}(b) looks qualitatively different from that in Fig.~\ref{Fig:sparaB}(a) as discussed in Ref.~\cite{Yoshihara17PRA}.
The parameters are obtained as
$\Delta/2\pi=1.68$~GHz, $\omega/2\pi=6.345$~GHz, and $g/2\pi=7.27$~GHz.
Here, $g$ is larger than both $\Delta$ and $\omega$, indicating that the circuit is in the deep-strong-coupling regime $[g \gtrsim \mathrm{max}(\omega,\sqrt{\Delta\omega}/2)]$~\cite{Ashhab10PRA,Casanova10PRL,GuKockum17PhysRep}.
The parameters from all the sets are summarized in Table~\ref{Tab:params}.

\begin{table}
\begin{tabular}{c c c c c c c}
    \hline
    \hline
\vspace{0.1cm}
\rule{0pt}{4ex} & ${\displaystyle \frac{\Delta}{2\pi}}$ & ${\displaystyle \frac{\omega}{2\pi}}$ & ${\displaystyle \frac{g}{2\pi}}$ & ${\displaystyle \frac{\Delta_0}{2\pi}}$ & ${\displaystyle \frac{\Delta_1}{2\pi}}$ & ${\displaystyle \frac{\Delta_2}{2\pi}}$ \\
    \hline
A & 1.246 & 6.365 & 0.42 & 1.236 & 1.215 & \\
&&&& (1.235) & (1.213)&\\
B & 1.01 & 6.296 & 5.41 & 0.233 & $-0.452$ & $-0.13$ \\
&&&& (0.229) & ($-0.448$) & ($-0.123$)\\
C & 0.92 & 6.288 & 5.59 & 0.193 & $-0.412$ & $-0.062$\\
&&&& (0.189) & ($-0.410$) & ($-0.059$)\\
D & 3.93 & 5.282 & 5.28 & 0.54 & $-1.512$ & 0.56\\
&&&& (0.539) & ($-1.503$) & (0.624)\\
E & 4.88 & 5.230 & 5.37 & 0.607 & $-1.746$ & 0.906\\
&&&& (0.607) & ($-1.741$) & (1.018)\\
F & 4.71 & 5.220 & 5.46 & 0.538 & $-1.642$ & 1.005\\
&&&& (0.542) & ($-1.641$) & (1.087)\\
G & 3.53 & 5.263 & 5.58 & 0.375 & $-1.255$ & 0.8\\
&&&& (0.379) & ($-1.244$) & (0.834)\\
H & 1.68 & 6.345 & 7.27 & 0.127 & $-0.518$ & 0.5\\
&&&& (0.122) & ($-0.514$) & (0.523)\\
I & 1.61 & 6.335 & 7.48 & 0.099 & $-0.458$ & 0.493\\
&&&& (0.099) & ($-0.451$) & (0.532)\\
    \hline
    \hline
\end{tabular}
\caption{Parameters of qubit-oscillator circuits in GHz.
$\Delta$, $\omega$, and $g$ are obtained from the (single-tone) transmission spectra.
The numbers for $\Delta_n$ ($n = 0$, 1, 2)
in the upper line for each data set are obtained from two-tone transmission spectra, while those in the lower line
(i.e. those between parentheses) are numerically calculated values using $\hat{H}_{\rm Rabi}$ and the parameters $\Delta$, $\omega$, and $g$.}
\label{Tab:params}
\end{table}

%Delta_n
To obtain the photon-number-dependent qubit frequency $\Delta_n$ ($n = 0$, 1, 2),
at least five transition frequencies out of seven allowed transitions [Fig.~\ref{Fig:CircuitElevels}(d)] are necessary.
However, in each spectrum at $\varepsilon = 0$,
we see only two signals at frequencies $\omega_{\textrm{g}0,\textrm{g}1}$ and $\omega_{\textrm{e}0,\textrm{e}1}$ corresponding, respectively, to the transitions $|\textrm{g}0\rangle \to |\textrm{g}1\rangle$ and $|\textrm{e}0\rangle \to |\textrm{e}1\rangle$,
which were also observed in our previous experiments~\cite{Yoshihara17NatPhys,Yoshihara17PRA}.
There are two main reasons behind this limitation on single-tone spectroscopy, where only a single-frequency weak probe signal is applied to the circuit.
First, only transition frequencies in the range of the measurement setup (in our case 4 to 8 GHz) can be measured.
Second, the signal from transitions that do not start from the lowest two energy levels will be weak because of the small thermal population of higher energy levels (in our case, the thermal population decreases by 2 orders of magnitude for each step up in the value of $n$).

%figure: two-tone
\begin{figure}
\includegraphics{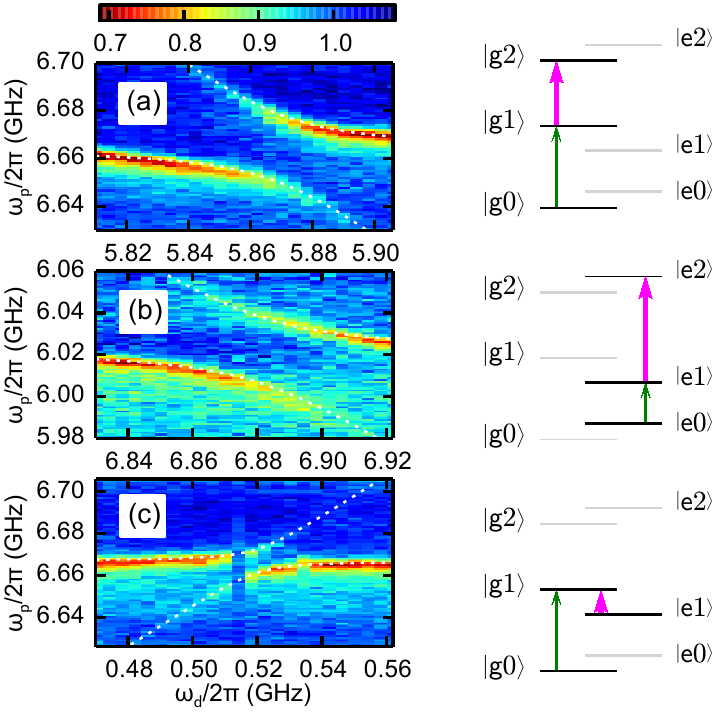}
\caption{(Left) Measured two-tone transmission spectra as functions of drive frequency $\omega_{\rm d} $ and probe frequency $\omega_{\rm p}$.
The color scheme is chosen such that the lowest point in each spectrum is red and the highest point is blue.
The white dotted lines are calculated transition frequencies considering dressed states due to
the drive signals.
The right panels show the energy-level diagrams.
The thin green arrows indicate transitions scanned by the probe signal,
while thick magenta arrows indicate transitions scanned by the drive signal.
}
\label{Fig:2tone}
\end{figure}
%two-tone spectroscopy
To access transitions other than $|\textrm{g}0\rangle \to |\textrm{g}1\rangle$ and $|\textrm{e}0\rangle \to |\textrm{e}1\rangle$, two-tone spectroscopy was used,
where a drive signal with frequency $\omega_{\rm d}$ is applied while the transmission of a probe signal with frequency $\omega_{\rm p}$ around the frequency $\omega_{\textrm{g}0,\textrm{g}1}$ or $\omega_{\textrm{e}0,\textrm{e}1}$ is measured.
When $\omega_{\rm d}$ is equal to the frequency of an allowed transition involving at least one of
the states $|\textrm{g}0\rangle$, $|\textrm{g}1\rangle$, $|\textrm{e}0\rangle$, and $|\textrm{e}1\rangle$,
an Autler-Townes splitting~\cite{Autler1955PR} takes place and is observed in the probe transmission signal.
Figure~\ref{Fig:2tone} shows the measured two-tone transmission spectra from set~H.
An avoided crossing between a horizontal line and a diagonal line~\cite{SM} is observed in each panel.
Interestingly, the slope of the diagonal line is $\partial \omega_{\rm p}/\partial \omega_{\rm d} = -1$ for Figs.~\ref{Fig:2tone}(a) and \ref{Fig:2tone}(b), and $+1$ for Fig.~\ref{Fig:2tone}(c),
which indicates that
the absorption of one probe photon is accompanied by the absorption of one photon from the drive field in Figs.~\ref{Fig:2tone}(a) and \ref{Fig:2tone}(b) and the emission of one photon to the drive field in Fig.~\ref{Fig:2tone}(c).
Together with the frequencies numerically calculated from $\hat{H}_{\rm Rabi}$, the corresponding transitions are identified as shown in the right-hand side of each spectrum.
The spectrum in Fig.~\ref{Fig:2tone}(c) demonstrates that the energy of $|\textrm{g}1\rangle$ is higher than that of $|\textrm{e}1\rangle$, and, hence, $\Delta_1$ is negative.
In other words, the qubit's energy levels are inverted.

%evaluate Delta_n
Moreover, from these three two-tone transmission spectra,
five transition frequencies, $\omega_{\textrm{g}0,\textrm{g}1}$,
$\omega_{\textrm{g}0,\textrm{g}2}$,
$\omega_{\textrm{e}0,\textrm{e}1}$,
$\omega_{\textrm{e}0,\textrm{e}2}$,
and
$\omega_{\textrm{g}0,\textrm{e}1}$,
can be evaluated.
In Fig.~\ref{Fig:2tone}(a), the horizontal line corresponds to a one-photon resonance, $\omega_{\rm p} = \omega_{\textrm{g}0,\textrm{g}1}$,
whereas the diagonal line corresponds to a two-photon resonance,
$\omega_{\rm p} = \omega_{\textrm{g}0,\textrm{g}2} -\omega_{\rm d}$.
For Fig.~\ref{Fig:2tone}(b) the horizontal line is at $\omega_{\rm p} = \omega_{\textrm{e}0,\textrm{e}1}$ and the diagonal line is
at $\omega_{\rm p} = \omega_{\textrm{e}0,\textrm{e}2} -\omega_{\rm d}$.
For Fig.~\ref{Fig:2tone}(c), the horizontal line is at
$\omega_{\rm p} = \omega_{\textrm{g}0,\textrm{g}1}$ and the diagonal line is 
at $\omega_{\rm p} = \omega_{\textrm{g}0,\textrm{e}1} +\omega_{\rm d}$.
From these five transition frequencies, all the eigenenergies up to the fifth excited state can be determined
up to an overall energy shift.
One thing is worth emphasizing here.
In the two-tone spectroscopy of a deep-strongly-coupled qubit-oscillator circuit,
the states of the qubit are doubly dressed:
one is the conventional dressing by the classical drive field,
while the other is in the quantum regime due to deep-strong coupling to the oscillator,
where the oscillator's states are displaced.
The experimental results demonstrate that the two kinds of dressed states coexist.

\begin{figure}
\includegraphics{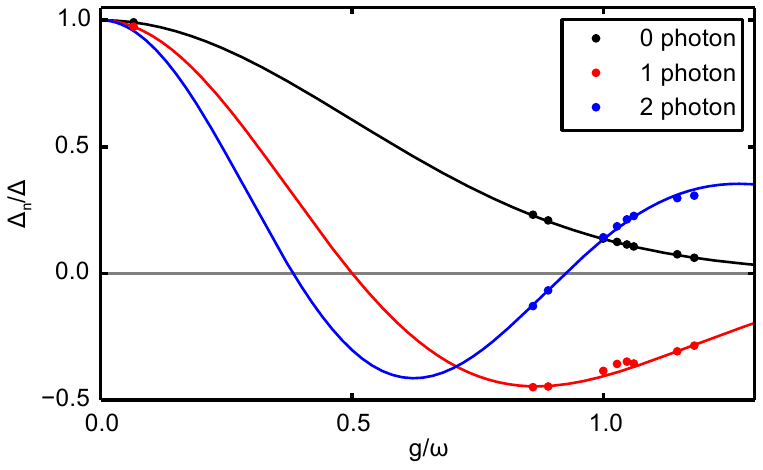}
\caption{Photon-number-dependent normalized qubit frequencies $\Delta_n/\Delta$ as functions of $g/\omega$.
The parameters $\Delta$, $\omega$, and $g$ are obtained from the single-tone transmission spectra.
The black, red, and blue solid circles are, respectively, the qubit frequencies $\Delta_0$, $\Delta_1$, and $\Delta_2$ obtained from the two-tone transmission spectra.
The solid lines are $\Delta_n$ obtained from Eq.~(\ref{Eq:Dn}).
}
\label{Fig:GauLag_both}
\end{figure}
%
%comparison between theory and experiment
From Eq.~(\ref{Eq:Dn}), the normalized photon-number-dependent qubit frequencies $\Delta_n/\Delta$ are expected to depend solely on the normalized coupling constant $g/\omega$.
We, therefore, plot $\Delta_n/\Delta$
as functions of $g/\omega$ for all nine parameter sets together (Fig.~\ref{Fig:GauLag_both}).
The parameters $\Delta$, $\omega$, and $g$ are obtained from the transmission spectra.
These results demonstrate huge Lamb shifts $|\Delta_0-\Delta|$,
some of them exceeding 90\% of the bare qubit frequencies $\Delta$.
These results also demonstrate that one-photon and two-photon ac-Stark shifts are so large that
$\Delta_1$ and $\Delta_2$ change their signs depending on $g/\omega$.
The solid lines are theoretically predicted values given by Eq.~(\ref{Eq:Dn}).
Table~\ref{Tab:params} shows a comparison between the measured and the numerically calculated $\Delta_n$~\cite{SM} using $\hat{H}_{\rm Rabi}$ and the parameters $\Delta$, $\omega$, and $g$.
In many circuits, the measured $\Delta_2$ is smaller than the numerically calculated one,
while the agreement of $\Delta_0$ and $\Delta_1$ is good,
with the deviations being at most 10~MHz.
Since $\Delta_2$ given by Eq.~(\ref{Eq:Dn}) is an approximation that becomes exact in the limit $\Delta/\omega \to 0$ while the numerically calculated $\Delta_2$ is based on the exact $\hat{H}_{\rm Rabi}$ for any set of parameters,
the agreement of $\Delta_2$ in Fig.~\ref{Fig:GauLag_both} is a coincidence.
In this way, our results can be used to check how well the flux qubit-LC oscillator circuits realize a system that is described by the quantum Rabi model Hamiltonian,
which is the basis for several important applications, e.g., ultrafast gates~\cite{Romero2012PRL} and quantum switches~\cite{Baust2016PRB}.
A possible source of the deviation in $\Delta_2$ is higher energy levels of the flux qubit.
As discussed in Ref.~\cite{Yoshihara17PRA},
the second or higher excited states can shift the energy levels of the qubit-oscillator circuit,
even though there is an energy difference of at least 20~GHz between the first and the second excited states.
Consideration of higher energy levels is necessary to identify the origin of the deviation in $\Delta_2$.

%conclusion
In conclusion, we have used two-tone spectroscopy to study
deep-strongly-coupled flux qubit-LC oscillator circuits.
We have determined the energies of the six lowest energy eigenstates of each circuit
and evaluated the photon-number-dependent qubit energy shifts.
We have found Lamb shifts that exceed 90\% of the bare qubit frequency
and inversions of the qubit's ground and excited states caused by the one-photon and two-photon ac-Stark shifts.
The results agree with the quantum Rabi model, giving further support to the validity of the quantum Rabi model in describing these circuits in the deep-strong-coupling regime.

%acknowledgment
\begin{acknowledgments}
We thank Masahiro Takeoka for stimulating discussions.
Z. A. acknowledges the Leading Graduate
Program in Science and Engineering, Waseda University from MEXT, Japan.
This work was supported by
Japan Society for the Promotion of Science
Grants-in-Aid for Scientific Research (KAKENHI) 
(Grants No. JP25220601 and No. JP16K05497), and 
Japan Science and Technology Agency 
Core Research for Evolutionary Science and Technology
(Grant No. JPMJCR1775).
\end{acknowledgments}

\section*{Supplemental Material}
\setcounter{figure}{0}
\setcounter{equation}{0}
\setcounter{table}{0}
\renewcommand\theequation{S\arabic{equation}}
\renewcommand\thesection{S\arabic{section}}
\renewcommand\thefigure{S\arabic{figure}}
\renewcommand\thetable{S\Roman{table}}
\section{overlap between the displaced Fock states}
\begin{figure}
\includegraphics{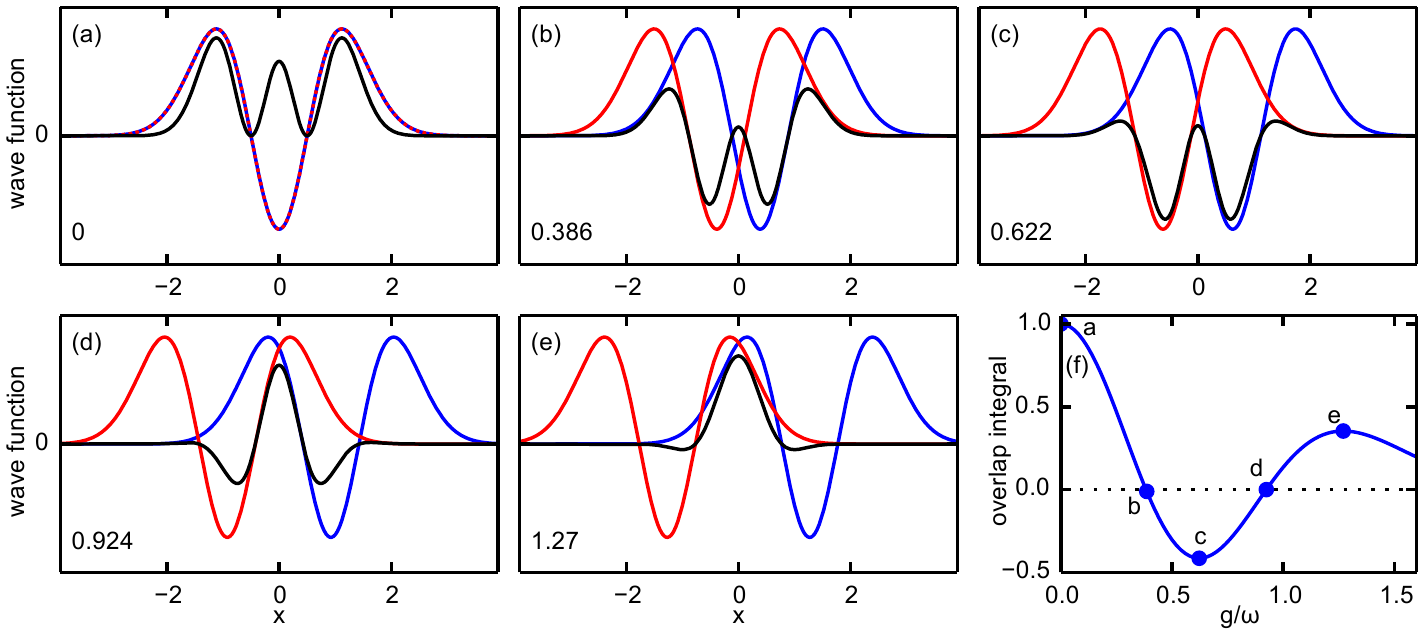}
\caption{(a)--(e) Wave functions of the oppositely displaced two-photon Fock state $\phi_2(x,-g/\omega)$ (red) and $\phi_2(x,g/\omega)$ (blue) and their product $\phi_2(x,-g/\omega)\phi_2(x,g/\omega)$ (black) for various values of $g/\omega$: (a)~0, (b)~0.386, (c)~0.622, (d)~0.924, and (e)~1.27. (f) Normalized overlap integral $I_2(g/\omega)/I_2(0)$. Solid blue circles indicate the values of $g/\omega$ for panels (a)--(e).
}
\label{Fig:He}
\end{figure}

In this section, we study the overlap between the two oppositely displaced Fock states,
$\hat{D}(g/\omega)|n\rangle_{\rm o}$ and $\hat{D}(-g/\omega)|n\rangle_{\rm o}$,
where $\hat{D}$ is the displacement operator,
$g$ is the coupling constant, $\omega$ is the resonance frequency of the oscillator,
and $|n\rangle_{\rm o}$ is the $n$-photon Fock state of the oscillator.
The wave function of $\hat{D}(g/\omega)|n\rangle_{\rm o}$ in the coordinate basis is given by
$\phi_n(x,g/\omega) = {}_{\rm o}\langle x|\hat{D}(g/\omega)|n\rangle_{\rm o}$,
where $|x\rangle_{\rm o}$ is an eigenstate of the coordinate operator
$\hat{x} = (\hat{a} + \hat{a}^\dagger)/2$,
and $\hat{a}$ and $\hat{a}^\dagger$ are the annihilation and creation operators.
Besides the normalization factor, it is given by
\begin{equation}
\phi_n(x,g/\omega) = \exp\{-[x-(g/\omega)]^2\}H_n(\sqrt{2}[x-(g/\omega)]),
\label{Eq:HOphi}
\end{equation}
where $H_n$ is the Hermite polynomial;
$H_0(x) = 1$, $H_1(x) = 2x$, $H_2(x) = 4x^2-2$, and so on.
Note that $\phi_n(x,g/\omega)$ is real.
The overlap integral, which appears in the second line of Eq.~(3) in the main text, can be calculated as
\begin{eqnarray}
I_n(g/\omega) & = & {}_{\rm o}\langle n | \hat{D}^\dagger (-g/\omega)\hat{D}(g/\omega)|n\rangle_{\rm o}\\
& = & \int_{-\infty}^{\infty} \phi_n(x,-g/\omega)\phi_n(x,g/\omega)dx.
\label{Eq:OLI}
\end{eqnarray}
To be concrete, in the following, we consider the case of $n = 2$ as an example.
Figures~\ref{Fig:He}(a)--(e) show wave functions of the displaced two-photon Fock states $\phi_2(x,\pm g/\omega)$ and their product $\phi_2(x,-g/\omega)\phi_2(x,g/\omega)$ for five different values of $g/\omega$.
The values of $g/\omega$ are chosen so that $I_2(g/\omega)$ is either maximal [(a) and (e)], minimal (c), or zero [(b) and (d)] [see Fig.~\ref{Fig:He}(f)].
When the positions of peaks or dips of $\phi_2(x,\pm g/\omega)$ coincide,
the product $\phi_2(x,-g/\omega)\phi_2(x,g/\omega)$ is mostly positive, and hence $I_2(g/\omega)$ becomes maximal [Figs.~\ref{Fig:He}(a) and (e)].
On the other hand, when the peak positions of a wave function coincide with the dip positions of the other,
the product $\phi_2(x,-g/\omega)\phi_2(x,g/\omega)$ is mostly negative, and hence $I_2(g/\omega)$ becomes minimal [Fig.~\ref{Fig:He}(c)].

\section{symmetry of quantum Rabi model and state assignment from the spectra}
The parity operator of the qubit-oscillator system is defined as $\hat{P} = \hat{P}_{\rm q}\otimes\hat{P}_{\rm o} \equiv \hat{\sigma}_z(-1)^{\hat{a}^\dagger \hat{a}}$,
where, $\hat{P}_{\rm q} \equiv \hat{\sigma}_z$ and $\hat{P}_{\rm o}\equiv (-1)^{\hat{a}^\dagger \hat{a}}$ are respectively the parity operators of the qubit and the oscillator.
The parities of states and operators are defined as follows;
The parity of a state $|\phi\rangle$ is even (odd) when $\hat{P}|\phi\rangle = |\phi\rangle$ $(-|\phi\rangle)$.
The parity of an operator $\hat{A}$ is
even when $[\hat{A},\hat{P}] = \hat{A}\hat{P} - \hat{P}\hat{A} = 0$, and is
odd when $\{\hat{A},\hat{P}\} = \hat{A}\hat{P} + \hat{P}\hat{A} = 0$.
The parity symmetry in the states and operators that appear in the quantum Rabi Hamiltonian,
\begin{equation}
\hat{H}_{\rm Rabi} = -\frac{\hbar\Delta}{2} \hat{\sigma}_z+\hbar\omega \hat{a}^{\dagger}\hat{a}+ \hbar g \hat{\sigma}_x \left( \hat{a} + \hat{a}^{\dagger} \right),
\label{Eq:RabiHami_SM}
\end{equation}
is summarized in Table~\ref{Tab:P}.
Here, $\Delta$ is the qubit's transition frequency.
Because both $\hat{\sigma}_x$ and $(\hat{a}+\hat{a}^\dagger)$ have negative
parities,
their product has a positive parity, meaning that all three terms in $\hat{H}_{\rm Rabi}$ have positive
parities.
Therefore, $[\hat{H}_{\rm Rabi}, \hat{P}] = 0$, and hence,
the energy eigenstates are also eigenstates of $\hat{P}$,
and have well-defined parities.
Note that this property does not depend on the values of $\Delta$, $\omega$, and $g$.

Although the energy eigenstates
and their eigenenegies
of $\hat{H}_{\rm Rabi}$ cannot be described
analytically
for arbitrary values of $\Delta$, $\omega$, and $g$,
the symmetry allows to define energy eigenstates
and their eigenenegies
of $\hat{H}_{\rm Rabi}$ as
$|in\rangle$
and $\omega_{in}$,
where $i$ (= g, e) indicates that the qubit is in ``g'' the ground or ``e'' the excited state and
$n$ is the number of real photons in the oscillator.
Since the parity of $(\hat{a} + \hat{a}^\dagger)$ is
odd,
the transition matrix elements $\langle im|(\hat{a} + \hat{a}^\dagger)|jn\rangle$
may have non-zero values
when the parities of the energy eigenstates $|im\rangle$ and $|jn\rangle$ are
opposite, whereas they always vanish
when the parities are same.

From the transition frequencies alone,
the energy eigenstates
and the eigenenergies
cannot be determined uniquely.
However, by using the parity symmetry discussed above,
energy eigenstates
and eigenenergies
are recursively determined as long as $\Delta < \omega$ in the following way.
(i) The ground and the first excited states of a coupled circuit are respectively $|\textrm{g}0\rangle$
and $|\textrm{e}0\rangle$, since there is no energy-level crossing between them.
The corresponding eigenenergies are respectively $\omega_{\textrm{g}0}$ and $\omega_{\textrm{e}0}$.
(ii) Between the (2$n$+2)th and $(2n + 3)$th excited states ($n \ge 0$), the state having nonzero transition matrix element with the state $|\textrm{g}n\rangle$ is $|\textrm{g}n+1\rangle$
and the other is $|\textrm{e}n+1\rangle$.
The corresponding eigenenergies are respectively $\omega_{\textrm{g}n+1}$ and $\omega_{\textrm{e}n+1}$. 
In this way, photon-number-dependent qubit frequency
$\Delta_n \equiv \omega_{\textrm{e}n} - \omega_{\textrm{g}n}$ can be uniquely
determined for all the parameter sets in this work.

\begin{table}
\begin{tabular}{l @{\hspace{0.2cm}}|@{\hspace{0.2cm}} c @{\hspace{0.5cm}} c}
    \hline
     parity & even & odd\\
%    \vspace{0.5cm}
    \hline
    \hline
\rule{0pt}{5ex}qubit state & ${\displaystyle |\textrm{g}\rangle_{\rm q} = \frac{\left |\circlearrowleft \right\rangle_{\rm q} + \left |\circlearrowright \right\rangle_{\rm q}}{\sqrt{2}}}$ & ${\displaystyle |\textrm{e}\rangle_{\rm q} = \frac{\left |\circlearrowleft \right\rangle_{\rm q} - \left |\circlearrowright \right\rangle_{\rm q}}{\sqrt{2}}}$\\
    photon state & $|$even number$\rangle_{\rm o}$ & $|$odd number$\rangle_{\rm o}$\\
    qubit operator & $\hat{\sigma}_z$ & $\hat{\sigma}_x$\\
    photon operator &$\hat{a}^\dagger\hat{a}$  & $\hat{a} + \hat{a}^\dagger$\\
    \hline
\end{tabular}
\caption{The parity symmetry in the states and operators that appears in $\hat{H}_{\rm Rabi}$.}
\label{Tab:P}
\end{table}

%section coupler inductance
\section{coupler inductance and flux bias points}
The coupler inductance for sets B--I is a dc superconducting quantum interference device (SQUID) consisting of two parallel Josephson junctions as shown in Fig.~1(c) in the main text.
Its Josephson inductance is given as
\begin{equation}
L_{\rm J} = \frac{\Phi_0}{2\pi \sqrt{(2I_{\rm c}\cos |\pi n_{\phi \rm c}|)^2-I_{\rm b}^2}},
\label{Eq:Ic_coupler}
\end{equation}
where $I_{\rm c}$ is the critical current of each Josephson junction,
$n_{\phi \rm c}$ is the normalized flux bias through the loop in units of the superconducting flux quantum $\Phi_0 = h/(2e)$,
and $I_{\rm b}$ is the current flowing through the SQUID.
An external superconducting magnet produces a uniform magnetic field,
and flux biases are applied to the qubit and the coupler proportionally to the areas of their loops.
The area ratio of the loops $r_{\rm c} = A_{\rm coupler}/A_{\rm qubit}$ is approximately 0.05.
The flux bias through the coupler loop $n_{\phi \rm c} = r_{\rm c}n_{\phi}$ depends on the normalized flux bias of the qubit $n_{\phi}$,
which in most cases is around the symmetry point of the qubit, i.e. $n_{\phi} = \pm 0.5, \pm 1.5$, and so on.

%section background transmission coefficient
\section{background transmission coefficient}
The amplitudes of the measured transmission spectra $|S_{21}^{\rm meas}(\varepsilon,\omega_{\rm p})|$ are fitted by the following formula:
\begin{eqnarray}
|S_{21}^{\rm meas}(\varepsilon,\omega_{\rm p})| = |S_{21}^{\rm bg}(\omega_{\rm p})S_{21}(\varepsilon,\omega_{\rm p})|,
\label{Eq:S21meas}
\end{eqnarray}
where 
\begin{eqnarray}
S_{21}(\varepsilon,\omega_{\rm p}) = 1-\frac{(Q_{\rm L}/Q_{\rm e})e^{i\phi}}{1+2iQ_{\rm L}\frac{\omega_{\rm p}-\omega_0(\varepsilon)}{\omega_0(\varepsilon)}},
\label{Eq:S21}
\end{eqnarray}
and we assumed that a background transmission coefficient $S_{21}^{\rm bg}(\omega_{\rm p})$ is independent of energy bias $\varepsilon$ and is written by a polynomial of the probe photon frequency $\omega_{\rm p}$.
Eq.~(\ref{Eq:S21}) can be applied to a transmission line that is inductively and capacitively coupled to an LC oscillator~\cite{Khalil12JAP},
where $Q_{\rm L}$ is the total quality factor of the oscillator,
$Q_{\rm e}$ is the external quality factor due to the coupling to the transmission line,
and $\phi$ is a phase factor that accounts for the asymmetry of the resonance line shape.
Note that $|S_{21}(\varepsilon,\omega_{\rm p})|$
may become larger than 1 depending on the value of $\phi$. 

%section avoided crossing
\section{avoided crossings in two-tone spectroscopy}
\begin{figure}
\includegraphics{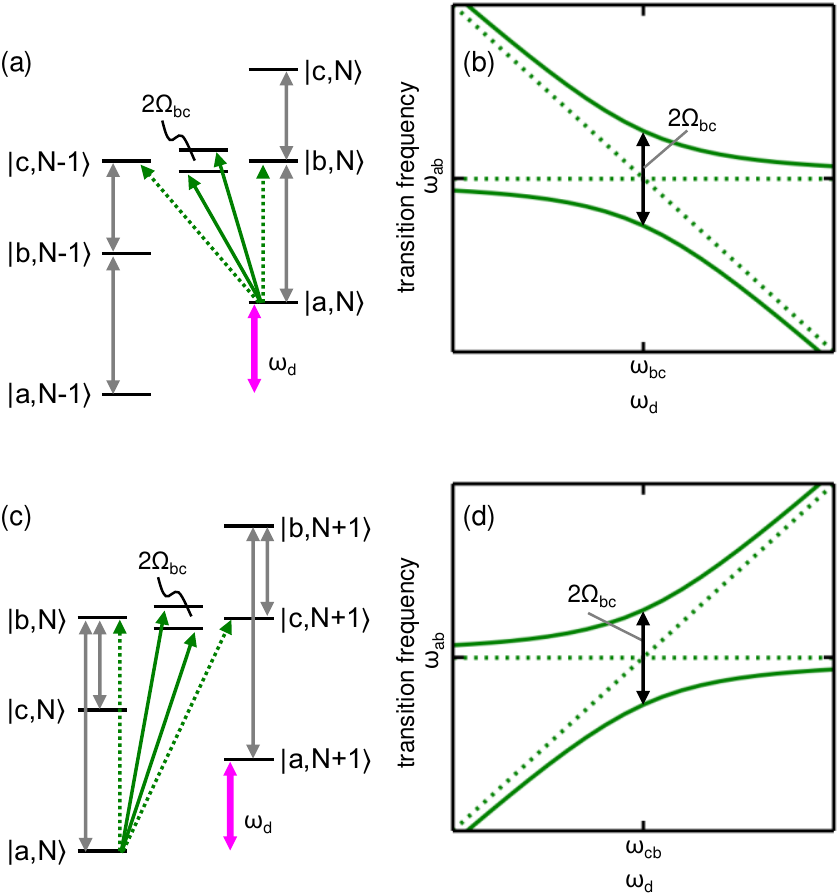}
\caption{(a) (c) The energy-level diagram when (a) $\omega_{\rm a} < \omega_{\rm b} < \omega_{\rm c}$ and (c) $\omega_{\rm a} < \omega_{\rm c} < \omega_{\rm b}$.
Here, the states $|i,k\rangle$ ($i$ = a, b, c, and $k$ is the number of drive photons) are energy eigenstates of $\mathcal{H}'_{\omega_{\rm b} < \omega_{\rm c}}$ and $\mathcal{H}'_{\omega_{\rm c} < \omega_{\rm b}}$ when the off-diagonal terms are ignored.
We have kept only the states with drive photon numbers $N-1$ and $N$ in (a) and $N$ and $N+1$ in (c).
The gray arrows indicate allowed transitions in the three-level system.
The magenta arrow indicates the drive frequency $\omega_{\rm d}$.
The dotted and solid green arrows indicate the transitions with the transition frequencies around $\omega_{\rm ab}$ when $\Omega_{\rm bc} = 0$ and $\Omega_{\rm bc} \neq 0$, respectively.
(b) (d) Transition frequencies as functions of $\omega_{\rm d}$ when (b) $\omega_{\rm a} < \omega_{\rm b} < \omega_{\rm c}$ and (d) $\omega_{\rm a} < \omega_{\rm c} < \omega_{\rm b}$.
The dotted and solid green lines correspond to the transition frequencies when $\Omega_{\rm bc} = 0$ and $\Omega_{\rm bc} \neq 0$, respectively.
}
\label{Fig:AC}
\end{figure}

In this section, we discuss the physical origin of the avoided crossings observed in Fig. 3 of the main text.
The Hamiltonian of a three-level system under the application of a drive field with frequency $\omega_{\rm d}$ can be described by
%Hamiltonian
\begin{eqnarray}
\nonumber
\mathcal{H}'_{\omega_{\rm b} < \omega_{\rm c}} & = & \hbar(\omega_{\rm a}\hat{\sigma}_{\rm aa} + \omega_{\rm b}\hat{\sigma}_{\rm bb} + \omega_{\rm c}\hat{\sigma}_{\rm cc})\\
&& + \hbar \omega_{\rm d}\hat{b}^\dagger\hat{b} + \hbar\left[\chi_{\rm ab}(\hat{\sigma}_{\rm ab}\hat{b}^\dagger + \hat{\sigma}_{\rm ba}\hat{b}) + \chi_{\rm bc}(\hat{\sigma}_{\rm bc}\hat{b}^\dagger + \hat{\sigma}_{\rm cb}\hat{b})\right],
\label{Eq:3lsbc}
\end{eqnarray}
where $\hat{\sigma}_{ij}=|i\rangle\langle j|$,
$\hat{b}$ and $\hat{b}^\dagger$ are respectively the annihilation and creation operators of the oscillator representing the drive field, and
$\chi_{ij}$ describes the interaction strength between the drive field mode and the transition dipole moment. Here we assume that $\omega_{\rm a} < \omega_{\rm b} < \omega_{\rm c}$ and the transition $|\mathrm{a}\rangle \to |\mathrm{c}\rangle$ is forbidden.
This situation applies to the energy eigenstates involved in the avoided crossings observed in Figs. 3(a) and (b) in the main text.
Namely, $|\mathrm{a}\rangle = |\mathrm{g}0\rangle$, $|\mathrm{b}\rangle = |\mathrm{g}1\rangle$,
and $|\mathrm{c}\rangle = |\mathrm{g}2\rangle$ for Fig. 3(a),
and $|\mathrm{a}\rangle = |\mathrm{e}0\rangle$, $|\mathrm{b}\rangle = |\mathrm{e}1\rangle$,
and $|\mathrm{c}\rangle = |\mathrm{e}2\rangle$ for Fig. 3(b).
The energy-level diagram of the coupled system (three-level system and drive field) is described in Fig.~\ref{Fig:AC}(a).
Here, the states $|i,k\rangle$ ($i$ = a, b, c, and $k$ is the number of drive photons) are energy eigenstates of $\mathcal{H}'_{\omega_{\rm b} < \omega_{\rm c}}$
when the off-diagonal terms are ignored.
The corresponding energies are given as $\omega_i + k\omega_{\rm d}$.
Under the application of the probe field, transitions occur between the energy eigenstates of $\mathcal{H}'_{\omega_{\rm b} < \omega_{\rm c}}$.
The $|\mathrm{a},N\rangle \to |\mathrm{b},N\rangle$ and $|\mathrm{a},N\rangle \to |\mathrm{c},N-1\rangle$ transition frequencies shown in Fig.~\ref{Fig:AC}(a) are given by $\omega_{\rm ab}$ and $\omega_{\rm ac} - \omega_{\rm d}$ ($\omega_{ij} = \omega_j - \omega_i$),
which are the horizontal and diagonal dotted lines in Fig.~\ref{Fig:AC}(b).
Note that the transition $|\mathrm{a},N\rangle \to |\mathrm{c},N-1\rangle$ accompanies the absorption of one photon from the drive field and hence the transition frequency decreases linearly with $\omega_{\rm d}$ with slope $-1$.
Note also that because the transitions $|\mathrm{a},N\rangle \to |\mathrm{c},N-1\rangle$ is forbidden when the drive field is off, the diagonal line does not appear in the spectrum.
The off diagonal terms of $\mathcal{H}'_{\omega_{\rm b} < \omega_{\rm c}}$ renormalize the energy eigenstates,
which are qubit-oscillator-drive doubly dressed states, and induces the avoided crossings.
In particular, for $\omega_{\rm d} \simeq \omega_{\rm bc}$,
we observe that $|\mathrm{b},N\rangle$ and $|\mathrm{c},N-1\rangle$ are nearly degenerate whereas the other states are largely separated from each other.
Then, the relevant eigenenergies of $\mathcal{H}'_{\omega_{\rm b} < \omega_{\rm c}}$ are approximately given by $\omega_{\rm a} + N\omega_{\rm d}$ and $[\omega_{\rm b} + \omega_{\rm c} + (2N - 1)\omega_{\rm d}]/2 \pm \sqrt{(\omega_{\rm bc}-\omega_{\rm d})^2/4 + \Omega^2_{\rm bc}}$,
where $\Omega_{ij} = \chi_{ij}\sqrt{N}$ and $N$ is the number of photons in the drive field.
The transition frequencies are then $(\omega_{\rm ab} + \omega_{\rm ac}-\omega_{\rm d})/2 \pm \sqrt{(\omega_{\rm bc}-\omega_{\rm d})^2/4 + \Omega^2_{\rm bc}}$, which are the solid lines in Fig.~\ref{Fig:AC}(b).

When $\omega_{\rm a} < \omega_{\rm c} < \omega_{\rm b}$ and $\chi_{\rm ac} = 0$,
the Hamiltonian is given by
\begin{eqnarray}
\nonumber
\mathcal{H}'_{\omega_{\rm c} < \omega_{\rm b}} & = & \hbar(\omega_{\rm a}\hat{\sigma}_{\rm aa} + \omega_{\rm b}\hat{\sigma}_{\rm bb} + \omega_{\rm c}\hat{\sigma}_{\rm cc})\\
&& + \hbar \omega_{\rm d}\hat{b}^\dagger\hat{b} + \hbar\left[\chi_{\rm ab}(\hat{\sigma}_{\rm ab}\hat{b}^\dagger + \hat{\sigma}_{\rm ba}\hat{b}) + \chi_{\rm bc}(\hat{\sigma}_{\rm cb}\hat{b}^\dagger + \hat{\sigma}_{\rm bc}\hat{b})\right].
\label{Eq:3lscb}
\end{eqnarray}
This situation applies to the energy eigenstates involved in the avoided crossing observed in Fig. 3(c) in the main text.
Namely, $|\mathrm{a}\rangle = |\mathrm{g}0\rangle$, $|\mathrm{b}\rangle = |\mathrm{g}1\rangle$,
and $|\mathrm{c}\rangle = |\mathrm{e}1\rangle$.
The energy-level diagram is described in Fig.~\ref{Fig:AC}(c).
The $|\mathrm{a},N\rangle \to |\mathrm{b},N\rangle$ and $|\mathrm{a},N\rangle \to |\mathrm{c},N + 1\rangle$ transition frequencies shown in Fig.~\ref{Fig:AC}(c) are given by $\omega_{\rm ab}$ and $\omega_{\rm ac} + \omega_{\rm d}$,
which are the horizontal and diagonal dotted lines in Fig.~\ref{Fig:AC}(d).
Note that the transition $|\mathrm{a},N\rangle \to |\mathrm{c},N+1\rangle$ accompanies the emission of one photon to the drive field and hence the transition frequency increases linearly with $\omega_{\rm d}$ with slope $+1$.
For $\omega_{\rm d} \simeq \omega_{\rm cb}$,
we observe that $|\mathrm{b},N\rangle$ and $|\mathrm{c},N+1\rangle$ are nearly degenerate whereas the other states are largely separated from each other.
The eigenenergies of $\mathcal{H}'_{\omega_{\rm c} < \omega_{\rm b}}$ are approximately given by $\omega_{\rm a} + N\omega_{\rm d}$ and $[\omega_{\rm b} + \omega_{\rm c} + (2N + 1)\omega_{\rm d}]/2 \pm \sqrt{(\omega_{\rm cb}-\omega_{\rm d})^2/4 + \Omega^2_{\rm bc}}$.
The transition frequencies are then $(\omega_{\rm ab} + \omega_{\rm ac}+\omega_{\rm d})/2 \pm \sqrt{(\omega_{\rm cb}-\omega_{\rm d})^2/4 + \Omega^2_{\rm bc}}$, which are the solid lines in Fig.~\ref{Fig:AC}(d).

%figure circuit diagram and energy levels

%numerically calculated $\Delta_n$
\section{numerically calculated $\Delta_n$}
\begin{figure}
\includegraphics{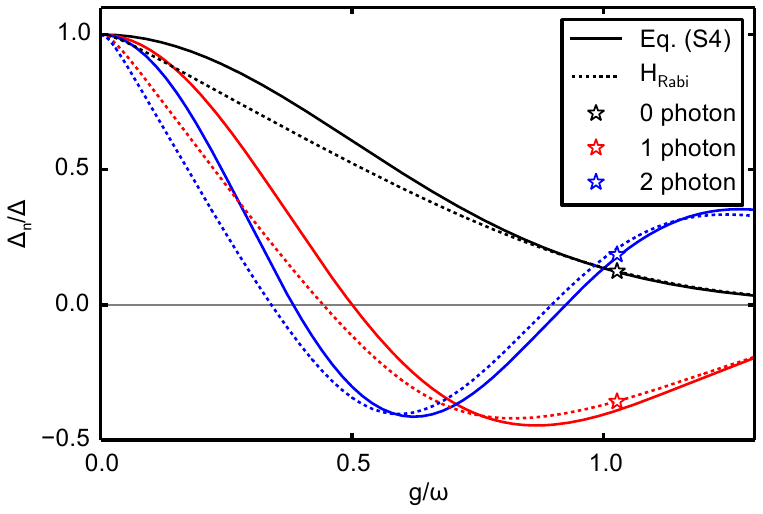}
\caption{Photon-number-dependent normalized qubit frequencies $\Delta_n/\Delta$ as functions of $g/\omega$.
The parameters $\Delta$, $\omega$, and $g$ are obtained from the transmission spectra.
The black, red, and blue solid lines are respectively $\Delta_0$, $\Delta_1$, and $\Delta_2$ obtained from Eq.~(\ref{Eq:Dn_SM}).
The dotted lines are numerically calculated $\Delta_n$ from $\hat{H}_{\rm Rabi}$ for $\Delta/\omega = 0.933$ corresponding to set~E.
The open stars are the qubit frequencies obtained from two-tone spectroscopies for set~E.
}
\label{Fig:GauLag_SM}
\end{figure}

In Fig.~\ref{Fig:GauLag_SM}, normalized photon-number-dependent qubit frequencies $\Delta_n/\Delta$ obtained from the two-tone spectroscopies are plotted in open stars for set~E,
which has largest value of $\Delta/\omega = 0.933$.
The solid lines are theoretically predicted values in the limit $\Delta \ll \omega$:
\begin{eqnarray}
\Delta_n(g/\omega) & \simeq & \Delta \exp(-2g^2/\omega^2)L_n(4g^2/\omega^2),
\label{Eq:Dn_SM}
\end{eqnarray}
which is also given in the main text.
The dotted lines are numerically calculated values from $\hat{H}_{\rm Rabi}$ and the parameter $\Delta/\omega = 0.933$.
Here, the eigenenergies and the energy eigenstates are calculated by diagonalizing $\hat{H}_{\rm Rabi}$,
where up to the 40-photon Fock states, which gives enough accuracy, are taken into account.
Some of our calculations were performed using the QuTiP simulation package~\cite{Johansson13CPC}.
Once we have the eigenenergies and the energy eigenstates,
$\Delta_n/\Delta$ can be obtained as discussed in Section~S2.
Note that the numerically calculated values approach the solid lines given by Eq.~(\ref{Eq:Dn_SM}) as the parameter $\Delta/\omega$ approaches zero.
Although there are clear deviations in smaller values of $g/\omega$,
the qualitative behaviors of solid and dotted lines are similar.
Unexpectedly, the blue open star (measured $\Delta_2$) is
close to the solid line rather than the dotted line,
although the latter is expected to give more accurate prediction
of the circuit described by $\hat{H}_{\rm Rabi}$.
The numerically calculated $\Delta_2$ in the range $0.8 \lesssim g/\omega \lesssim 1.1$ is larger than $\Delta_2$ given by Eq.~(\ref{Eq:Dn_SM}) and hence
the agreement of the blue open star and the solid line is a coincidence.

\end{document}